\documentclass[manuscript,screen,review=false]{acmart}
\AtBeginDocument{%
  \providecommand\BibTeX{{%
    \normalfont B\kern-0.5em{\scshape i\kern-0.25em b}\kern-0.8em\TeX}}}

\setcopyright{acmcopyright}
\copyrightyear{2022}
\acmYear{2022}

\acmConference[CHI 2022]{Make sure to enter the correct
  conference title from your rights confirmation emai}{SpaceCHI 2.0 workshop at ACM SIGCHI 2022}{New Orleans, LA}
\begin{document}
\makeatletter
\renewcommand\@formatdoi[1]{\ignorespaces}
\makeatother
\title{Challenges and Opportunities for Inclusive Space Missions}

\author{Markus Wieland}
\email{markus.wieland@ifi.lmu.de}
\affiliation{%
  \institution{LMU Munich}
  \country{Germany}
}

\author{Sebastian Feger}
\email{sebastian.feger@ifi.lmu.de}
\affiliation{%
  \institution{LMU Munich}
  \country{Germany}
}

\author{Tonja Machulla}
\email{tonja.machulla@ifi.lmu.de}
\affiliation{%
  \institution{TU Dortmund}
  \country{Germany}
}

\author{Albrecht Schmidt}
\email{albrecht.schmidt@ifi.lmu.de}
\affiliation{%
  \institution{LMU Munich}
  \country{Germany}
}

\renewcommand{\shortauthors}{Wieland et al.}

\begin{abstract}
People with impairments are not able to participate in space missions. However, this is not because they cannot; instead, spacecraft have not been designed for them. Therefore, instead of simply excluding people with impairments, they should be included in the design process. This may also help astronauts' suffering from fatigue or accidents. Indeed, several impairments can occur due to the effects of microgravity on the body and psychological factors during long space missions. In this paper, we describe the idea of including people with all types of impairments in the design process of interactive space systems to obtain, as a result, systems that can also be used by astronauts' suffering from a temporary or situational impairment during a long space flight. To this end, we have described solutions for some types of impairments to ensure the use of interactive systems with permanent, temporary, or situational impairments. The benefits from the participation of people with impairments also bring the idea of inclusion in space a bit closer: supporting people with impairments in space by designing appropriate systems that make space accessible.
\end{abstract}


\maketitle

\section{Introduction}
In the call for applications for the 2021 astronaut program, the European Space Agency (ESA) invited people with disabilities to apply for the first time. In addition, ESA has launched the Parastronaut Feasibility Project, which aims to identify the challenges people with specific disabilities may face on a space mission \cite{esa_parastronaut_2021}.
However, applications were not open to all types of disabilities. The call lists three types of disabilities: lower limb deficiency, leg difference, and individuals with less than 130cm height.
Regardless, this is a seminal step, hopefully followed by further efforts with regard to the inclusion of other types of disabilities \cite{heinicke_disability_2021}. 

An open application process and systems tailored to people with impairments would also be a step towards an inclusive space, as disability is defined as long-term physical, mental, intellectual, or sensory impairments in interaction with various barriers \cite{united_nations_convention_2006}. Thus, challenges for people with impairment arise through barriers in their environment. The consideration of all types of disabilities not only includes people with impairments and thus reduces barriers, but also opens up possibilities in the design of interactive systems, since temporary or situational disabilities may also occur during longer missions in space.

In this paper, we describe potential applications of interactive systems that particularly benefit from including people with various forms of disabilities in the design process, as impairments in space may be temporary or situational. For example, the astronauts'
performance is critical to mission success, and it is suggested that cognitive performance declines during extended missions in space \cite{casler_cognitive_1999,roy-oreilly_review_2021}. Further, a well-known fact is the change and thus slight loss of vision during and after long missions in space and is called spaceflight associated neuro-ocular syndrome (SANS) \cite{lee_spaceflight_2020}. SANS includes various ocular impairments such as unilateral and bilateral optic disc edema, globe flattening, choroidal and retinal folds, and more \cite{lee_space_2018}. Additionally, traumatic injuries during space missions can occur and affect astronauts' in their daily work and in the operation of interactive systems. The most common are hand injuries during space missions \cite{scheuring_musculoskeletal_2009}. Therefore, including people with all types of disabilities in the design process of interactive systems can lead to benefits for every astronaut on space missions.

Accordingly, we outline potential use cases for some types of disabilities and how these disabilities may occur temporarily during extended space missions. We also emphasize the importance of designing and developing interactive systems that all people can use.

\section{Challenges for inclusive Design}
Here, we describe possible challenges for interactive systems for selected impairments from the general categories of cognitive, sensory, and motor disabilities.

\subsection{Design Challenges for Cognitive Impairments}
We refer to external effects that can impact cognitive performance by cognitive impairment. Here, two different effects can influence cognitive performance in space: microgravity and effects related to workload, sleep, and physical or emotional facts on the astronauts' \cite{kanas_human_2008}. First, long-duration spaceflights and the associated microgravity are thought to cause changes in brain structure and therefore affect motor and cognitive functions \cite{roberts_prolonged_2019}. Further, even during long-duration spaceflight, cognitive performance was lower compared to before and after the space mission \cite{takacs_persistent_2021}. Second, the extreme conditions for long-duration spaceflight can induce various types of stress in astronauts' \cite{kanas_human_2008}. It is well known that stress affects cognitive performance, which has been demonstrated in earth-based research \cite{albery_effect_1989}. Thus, there are two types of influences on astronauts' cognitive performance during prolonged spaceflights. However, there are interindividual differences in cognitive performance across studies, which may be due to the small sample size \cite{strangman_human_2014}.

When these two types of influences impair cognitive performance, one human-computer interaction challenge is to design interactive systems to respond appropriately to reduce cognitive performance. Therefore, the interactive system should be able to measure the current cognitive load of an astronaut (e.g., through pupil dilation \cite{zagermann_measuring_2016,unsworth_tracking_2018}) and have knowledge of the overall goal of the task in order to provide appropriate context-specific support. Thus, the interactive systems could adapt content specifically to the current situation. In addition, testing the interactive system with people with cognitive impairments is crucial.

\subsection{Design Challenges for Sensory Impairments}
Since visual impairments could become a problem during extended space missions and nearly 50\% of astronauts' on long-term space missions had deterioration of their distant and near visual acuity, it is even more important to provide potential support for astronauts' \cite{mader_optic_2011}. A functional magnetic resonance imaging (fMRI) study compared two groups of astronauts' who spent a short and long time in space and showed that the astronauts' group that spent more time in space had more cerebrospinal fluid (CSF) in their brains than the short-trip astronauts', which exert pressure on the eyeball, flattening it and thereby displacing the optic nerve \cite{heldt_spaceflight-induced_2018}.

When spaceflight-associated neuro-ocular syndrome (SANS) occurs in an astronaut and visual performance deteriorates, the astronaut must still be able to operate interactive systems. Assistive technologies can provide support, such as those already used for people with visual impairments. Smartglasses are used for people with visual impairments, for example, to facilitate navigation \cite{zhao_designing_2019}, to enhance edges and magnify objects such as a wristwatch \cite{zhao_foresee_2015}, and for highlighting elements on a touch user interface \cite{lang_pressing_2021}. The existing approaches for visually impaired people can be evaluated for space missions and possibly used with an astronaut with SANS.

By including deaf and hard-of-hearing people, critical alarms and warnings, which use audio, need to be redesigned for visual feedback. This offers new perspectives, as excessive continuous noise levels (e.g., from spacecraft machinery) can have adverse effects on human health and performance \cite{limardo_international_2017}. Deaf and hard-of-hearing people might have an advantage here, as they are not much exposed to the continuous noise. However, the other astronauts' may learn sign language to communicate with deaf people.

Combining and connecting different research areas (e.g., aerospace engineering and assistive technologies for people with sensory impairments) could benefit astronauts' and help them in their daily work. Further research is needed to evaluate assistive technologies in harsh space environments.

\subsection{Design Challenges for Motor Impairments}
Hand injuries often occur in space as abrasions and cuts \cite{scheuring_musculoskeletal_2009}. Further, astronauts' training for an extra-vehicular activity often face the problem of the pressure in their spacesuit gloves, creating pressure points that can lead to pain, muscle fatigue, and even detaching the fingernail from the nail bed \cite{charvat_spacesuit_2015}. It may also happen that, situationally one hand of the astronaut cannot be used to operate the interactive system. This can affect the operation of an interactive system, especially when two hands are needed simultaneously. To address this challenge, smartglasses were tested during a task as a hands-free interaction tool, with participants using voice commands and additional information displayed through the lenses for task-specific requirements \cite{helin_mixed_2019}. In addition, various voice user interfaces (VUI) for extraterrestrial habitats have been tested to provide mission-specific assistance during scientific experiments \cite{goncalves_freitas_conversational_2021}.

In general, VUIs are promising and suitable for use on long space missions because they do not require hands or arms. Furthermore, the voice could also be used to control various devices or vehicles. Therefore, the design of interactive systems should take into account situational and temporary motor impairments that may occur during an extended space mission and incorporate VUIs to facilitate task completion for people with and without impairments.

\section{Conclusion}
We presented how the inclusion of people with impairments can positively impact the design and thus the use of interactive systems in space. In addition, we have described that the impairments mentioned above can also occur during long space missions. We divided the term disability into cognitive, sensory, and motor and proposed different approaches. The impairments and ideas presented are intended to provide a brief overview and are far from complete and serve as a starting point. More importantly, we would like to emphasize the importance of aerospace engineers and assistive technology researchers working together to move closer to the idea of an inclusive space that benefits everyone. Maybe in the future, it will no longer require specific calls for astronauts' with disabilities; instead, their application will simply be a given. Consequently, a first step could be taken in the direction of inclusive space: design spacecraft to support people with impairments and make space exploration accessible.




\newpage
\bibliographystyle{ACM-Reference-Format}
\bibliography{Chi}


\end{document}